\documentclass[journal]{IEEEtran} 				
\IEEEoverridecommandlockouts 						
\usepackage{amsmath,amssymb}
\usepackage{amsthm}
\usepackage{color}
\usepackage[mathscr]{eucal}
\usepackage{graphics,graphicx,multicol}
\usepackage{epstopdf}
\usepackage{enumerate}
\usepackage{subfigure}
\usepackage{algorithmic}
\usepackage{algorithm}
\usepackage{morefloats}
\usepackage{enumerate}
\usepackage{subfigure}
\usepackage{multirow}
\usepackage{array}
\usepackage{pstricks, pst-node, pst-plot, pst-circ}
\usepackage{moredefs}
\usepackage{courier}
\usepackage{mathtools}

\begin{document}

\title{Multi-Application Resource Allocation with Users Discrimination in Cellular Networks}
\author{Haya Shajaiah, Ahmed Abdelhadi and Charles Clancy \\
Hume Center, Virginia Tech, Arlington, VA, 22203, USA\\
\{hayajs, aabdelhadi, tcc\}@vt.edu
}
\maketitle

\begin{abstract}
In this paper, we consider resource allocation optimization problem in cellular networks for different types of users running multiple applications simultaneously. In our proposed model, each user application is assigned a utility function that represents the application type running on the user equipment (UE). The network operators assign a subscription weight to each UE based on its subscription. Each UE assigns an application weight to each of its applications based on the instantaneous usage percentage of the application. Additionally, UEs with higher priority assign applications target rates to their applications. Our objective is to allocate the resources optimally among the UEs and their applications from a single evolved node B (eNodeB) based on a utility proportional fairness policy with priority to real-time application users. A minimum quality of service (QoS) is guaranteed to each UE application based on the UE subscription weight, the UE application weight and the UE application target rate.  We propose a two-stage rate allocation algorithm to allocate the eNodeB resources among users and their applications. 
Finally, we present simulation results for the performance of our rate allocation algorithm.
\end{abstract}


\providelength{\AxesLineWidth}       \setlength{\AxesLineWidth}{0.5pt}%
\providelength{\plotwidth}           \setlength{\plotwidth}{8cm}
\providelength{\LineWidth}           \setlength{\LineWidth}{0.7pt}%
\providelength{\MarkerSize}          \setlength{\MarkerSize}{3pt}%
\newrgbcolor{GridColor}{0.8 0.8 0.8}%
\newrgbcolor{GridColor2}{0.5 0.5 0.5}%

\section{Introduction}\label{sec:intro}

The number of smart phones users and their traffic are increasing rapidly in recent years. Mobile users are now running multiple applications simultaneously on their smart phones. 
Operators are moving from single-service to multi-service and new services such as multimedia telephony and mobile-TV are now provided. The mobile users applications require different bit rates and packet delays based on the nature of the application. 
In addition, different users subscribing for the same service may receive different treatment from the network providers \cite{QoScontrol} because of the subscriber differentiation provided by the service providers.

A bandwidth proportional fairness algorithm (Frank Kelly algorithm) is introduced in \cite{kelly98ratecontrol}. The algorithms at the links are based on Lagrange multiplier methods of optimization theory, so the concavity assumption is satisfied. 
Sigmoidal-like utility functions with different parameters are used to represent real-time applications such as voice-over-IP (VoIP) and streaming video in \cite{Fundamental}.
Network utility optimization is used in \cite{RebeccaThesis} and \cite{DL_PowerAllocation} to allocate resources for real-time applications. A utility proportional fairness resource allocation approach is introduced in \cite{Ahmed_Utility1}, where the objective is to provide fair percentage on utility for each user. Different types of users traffic are considered. 
The authors have proven that the optimization problem is a convex optimization problem and therefore a tractable global optimal solution exists. Furthermore, a distributed rate allocation algorithm is presented by the authors to allocate the eNodeB resources optimally among users.

In this paper, we focus on finding an optimal solution for the resource allocation problem for different types of users running multiple types of applications simultaneously on their UEs. 
Each user can run multiple applications simultaneously and each application is represented by a utility function based on the application type. In addition, each application is assigned an application weight by the UE based on the application instantaneous usage percentage and importance to the UE. 
Furthermore, certain type of users with higher priority (e.g. VIP users) are assigned applications target rates by the network. Therefore, these VIP UEs' applications are given higher priority by the network when allocating resources. A minimum QoS is guaranteed for each user by using a proportional fairness approach and real-time applications are given priority over delay-tolerant applications.
Our resource allocation algorithm is performed in two stages. In the first stage, the eNodeB collaborates with the UEs to allocate user rates. In the second stage, the rates are allocated to user applications internally by the UEs.
\subsection{Related Work}\label{sec:related}
In \cite{Ahmed_Utility1}, \cite{Ahmed_Utility2} and \cite{Ahmed_Utility3}, the authors present an optimal rate allocation algorithm for users connected to a single carrier. The optimal rates are achieved by formulating the rate allocation optimization problem in a convex optimization framework. 
In \cite{Ahmed_Utility3}, the authors considered a resource allocation optimization problem with service-offering differentiation and application status differentiation. In their system model, each subscriber is running multiple applications simultaneously. 
The rate allocation algorithm is achieved in two stages.
%

A rate allocation with carrier aggregation approach is presented in \cite{Haya_Utility1}, the authors used two stage algorithm to allocate two carriers resources optimally among UEs. 
In \cite{Haya_Utility2}, the authors present a resource allocation optimization problem for two groups of users. 
The two groups are commercial and public safety users. The algorithm gives priority to the public safety users when allocating the eNodeB resources.
\subsection{Our Contributions}\label{sec:contributions}
Our contributions in this paper are summarized as:
\begin{itemize}
\item We present a resource allocation optimization problem to allocate the eNodeB resources optimally among different types of users running multiple applications.
\item We propose a two-stage rate allocation method to allocate rates optimally among users. First, the eNodeB and the UE collaborate to allocate an optimal rate to each UE. Each UE then allocates its assigned rate optimally among its applications.
\item We show that our resource allocation optimization problems have unique tractable global optimal solutions.
\end{itemize}
The remainder of this paper is organized as follows. Section \ref{sec:Problem_formulation} presents the problem formulation. In section \ref{sec:ResourceAllocation}, we present the two cases resource allocation optimization problems. Section \ref{sec:two-stage-alg} presents our two-stage rate allocation algorithm for the utility proportional fairness policy. In section \ref{sec:sim}, we discuss simulation setup and provide quantitative results along with discussion. Section \ref{sec:conclude} concludes the paper.
\section{Problem Formulation}\label{sec:Problem_formulation}
We consider a single cell mobile system that consists of a single eNodeB, $M$ regular UEs and another $N$ VIP UEs as shown in Figure \ref{fig:system_model1}. The rate allocated by the eNodeB to the $i^{th}$ UE is given by $r_i$. Each UE has its own utility function $X_i(r_i)$ that corresponds to the user satisfaction with its allocated rate $r_i$. Our objective is first to determine the optimal rates the eNodeB shall allocate to the UEs. We assume that the utility function $X_i(r_i)$ that is assigned to the $i^{th}$ user is given by:
\begin{figure}
 \includegraphics[height=1.7in, width=3.2in]{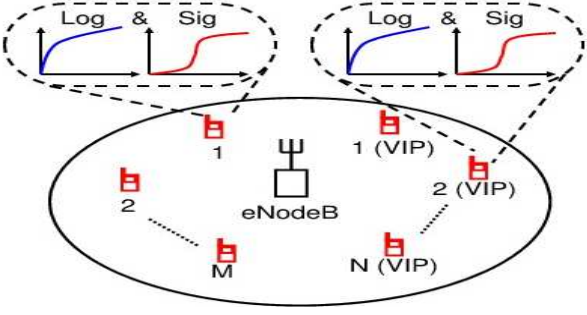}
 \caption{System Model, one eNodeB with $N$ VIP UEs and another $M$ regular UEs subscribing for a mobile service in the eNodeB coverage area.}
 \label{fig:system_model1}
 \end{figure}

\begin{equation}\label{eqn:utility_agg}
X_i(r_i) = \prod_{j=1}^{L_i}U_{ij}^{\alpha_{ij}}(r_{ij}+c_{ij})
\end{equation}
\[
  c_{ij} = \begin{dcases*}
         r_{ij}^{t} & if the $j^{th}$ application is assigned\\
            & an application target rate\\
         0 & if the $j^{th}$ application is not assigned\\
            & an application target rate
         \end{dcases*}
 \]
where $U_{ij}(r_{ij})$ is the $j^{th}$ application utility function for user $i$, $r_{ij}$ is the rate allocated to the $j^{th}$ application running on the $i^{th}$ UE, $L_i$ is the number of applications running on the $i^{th}$ UE, $c_{ij}$ is the application target rate for the $j^{th}$ application of user $i$ if it is assigned one whereas it is $0$ if the $j^{th}$ application is not assigned an application target rate by the network, $\alpha_{ij}$ is the $j^{th}$ application usage percentage (application weight) of the $i^{th}$ UE and $r_{ij}^{t}$ is the application target rate assigned to the $j^{th}$ application of the $i^{th}$ user. 

We express the user satisfaction with its provided service using utility functions \cite{DL_PowerAllocation} \cite{Fundamental} \cite{Utility-proportional}. We assume that the $j^{th}$ application utility function for user $i$ is given by $U_{ij} (r_{ij})$ that is strictly concave or sigmoidal-like function where $r_{ij}$ is the rate allocated to the $j^{th}$ application of user $i$. Delay tolerant applications are represented by logarithmic utility functions whereas real-time applications are represented by sigmoidal-like utility functions. These utility functions have the following properties:
\begin{itemize}
\item $U_{ij}(0) = 0$ and $U_{ij}(r_{ij})$ is an increasing function of $r_{ij}$.
\item $U_{ij}(r_{ij})$ is twice continuously differentiable in $r_{ij}$ and bounded above.
\end{itemize}
In our model, we use the normalized sigmoidal-like utility function, as in \cite{DL_PowerAllocation}, that can be expressed as
\begin{equation}\label{eqn:sigmoid}
U_{ij}(r_{ij}) = c_{ij}\Big(\frac{1}{1+e^{-a_{ij}(r_{ij}-b_{ij})}}-d_{ij}\Big)
\end{equation}
where $c_{ij} = \frac{1+e^{a_{ij}b_{ij}}}{e^{a_{ij}b_{ij}}}$ and $d_{ij} = \frac{1}{1+e^{a_{ij}b_{ij}}}$. So, it satisfies $U_{ij}(0)=0$ and $U_{ij}(\infty)=1$. The inflection point of normalized sigmoidal-like function is at $r_{ij}^{\text{inf}}=b_{ij}$.
In addition, we use the normalized logarithmic utility function, as in \cite{Ahmed_Utility1}, that can be expressed as
\begin{equation}\label{eqn:log}
U_{ij}(r_{ij}) = \frac{\log(1+k_{ij}r_{ij})}{\log(1+k_{ij}r_{max})}
\end{equation}
where $r_{max}$ is the maximum required rate for the user application to achieve 100\% utilization and $k_{ij}$ is the rate of increase of utilization with the allocated rate $r_{ij}$. So, it satisfies $U_{ij}(0)=0$ and $U_{ij}(r_{max})=1$. The inflection point of normalized logarithmic function is at $r_{ij}^{\text{inf}}=0$.
\section{Resource Allocation Optimization Problem}\label{sec:ResourceAllocation}
The resource allocation (RA) optimization problem for multi-application users is divided into two cases. The first-case is when the maximum available resources $R$ of the eNodeB is less than or equal to the total VIP UEs applications target rates. The second-case is when $R$ is greater than the total UEs applications target rates. The RA optimization problems for the two cases will be solved by our proposed algorithm to obtain the optimal rate for each UE as well as the optimal rates for the UE applications.

\subsection{First-Case RA Optimization Problem when $\sum_{i=1}^{M}\sum_{j=1}^{L_i}r_{ij}^t\geq R$} \label{First-Case-RA}
In this case, the eNodeB only allocates resources to the $M$ VIP UEs as they are considered more important and regular users will not be allocated any of the eNodeB resources since its available resources are limited. In this case, the optimization problem is divided into two stages. In the first-stage, the eNodeB allocates rates $r_i$ to the $M$ group of users. Both the eNodeB and the $M$ UEs collaborate to achieve the UEs resource allocation. In the second-stage, each one of these $M$ UEs uses the rate allocated to it by the eNodeB to allocate optimal rates $r_{ij}$ to its  $L_i$ applications. The second-stage is performed internally in the UE.
\subsubsection{First-Stage of the First-Case Optimization Problem}\label{FC-FS-RA}
In this case, the optimization problem for the first-stage can be written as:
\begin{equation}\label{eqn:opt_sub1}
\begin{aligned}
& \underset{\textbf{r}}{\text{max}}
& & \prod_{i=1}^{M}X_i^{\beta_{i}}(r_i) \\
& \text{subject to}
& & \sum_{i=1}^{M}r_i \leq R\\
& & & 0 \leq  r_i \leq \sum_{j=1}^{L_i}r_{ij}^t, \;\;\;\;\; i = 1,2, ...,M.
\end{aligned}
\end{equation}
where $X_i=\prod_{j=1}^{L_i}U_{ij}^{\alpha_{ij}}(r_{ij})$, $\textbf{r} =\{r_1,r_2,...,r_M\}$ is the rate allocated by the eNodeB to the $i^{th}$ UE, $M$ is the number of VIP UEs in the coverage area of the eNodeB, $R$ is the maximum achievable rate of the given eNodeB and $\beta_i$ is the $i^{th}$ user subscription weight assigned by the network.

The objective function in the optimization problem (\ref{eqn:opt_sub1}) is equivalent to $\sum_{i=1}^{M}\beta_i \log(X_i(r_i))$. Therefore, the optimization problem (\ref{eqn:opt_sub1}) is a convex optimization problem and there exists a unique tractable global optimal solution as shown in Corollary (III.1) \cite{Ahmed_Utility3}. This optimal solution gives each of the $M$ users an optimal rate $r_i^{\text{opt}}$ that is  less than or equal to the total applications target rates for that UE.
%
%
%
%
%
%
\subsubsection{Second-Stage of the First-Case Optimization Problem}\label{FC-SS-RA}
Each one of the $M$ VIP UEs allocates optimal rates $r_{ij}^{opt}$ to its $L_i$ applications. The optimal rate allocated to each application depends on the application differentiation weight and the application type. This optimization problem is solved internally in the UE and can be written for the $i^{th}$ UE as follows:
\begin{equation}\label{eqn:opt_sub2}
\begin{aligned}
& \underset{\textbf{r}_i}{\text{max}}
& & \prod_{j=1}^{L_i}U_{ij}^{\alpha_{ij}}(r_{ij}) \\
& \text{subject to}
& & \sum_{j=1}^{L_i}r_{ij} \leq r_i^{\text{opt}}\\
& & & 0 \leq r_{ij} \leq r_{ij}^{\text{t}}, \;\;\;\;\; j = 1,2, ...,L_i.
\end{aligned}
\end{equation}
where $\textbf{r}_i =\{r_{i1},r_{i2},...,r_{iL_i}\}$, $r_i^{\text{opt}}$ is the optimal rate allocated by the eNodeB to the $i^{th}$ UE and $L_i$ is number of the UE applications.
Since the objective function in the optimization problem (\ref{eqn:opt_sub2}) is equivalent to $\sum_{j=1}^{L_i}\alpha_{ij} \log(U_{ij}(r_{ij}))$,
%
then optimization problem (\ref{eqn:opt_sub2}) is convex and there exists a unique tractable global optimal solution as shown in Corollary (III.2) \cite{Ahmed_Utility3}. This optimal solution represents the optimal rate $r_{ij}^{\text{opt}}$ allocated to each of the $L_i$ applications.
%
%
%
%
\subsection{Second-Case RA Optimization Problem when $\sum_{i=1}^{M}\sum_{j=1}^{L_i}r_{ij}^t\textless R$} \label{Second-Case-RA}
In this case, the eNodeB first allocates resources to the $M$ VIP UEs. It then allocates the remaining resources based on the proportional fairness approach. The optimization problem in this case is divided into two stages. In the first-stage, the eNodeB collaborates with the UEs to allocate rates $r_i$ to all UEs. In the second-stage, each one of these $M+N$ UEs allocates optimal rates $r_{ij}$ to its applications. The second-stage is performed internally in the UE. The inelastic traffic are given priority when allocating the resources internally by the UEs.
\subsubsection{First-Stage of the Second-Case Optimization Problem}\label{FC-SS-RA}
%
In this case, the optimization problem of the first-stage can be written as:
\begin{equation}\label{eqn:SC_opt_sub1}
\begin{aligned}
& \underset{\textbf{r}}{\text{max}}
& & \prod_{i=1}^{M+N}X_i^{\beta_{i}}(r_i) \\
& \text{subject to}
& & \sum_{i=1}^{M+N}r_i \leq R\\
& & & r_i \geq 0, \;\;\;\;\; i = 1,2,...,M+N.
\end{aligned}
\end{equation}
where $X_i=\prod_{j=1}^{L_i}U_{ij}^{\alpha_{ij}}(r_{ij}+c_{ij})$ and $\textbf{r} =\{r_1,r_2,...,r_{M+N}\}$ and $M+N$ is the number of the VIP and regular UEs subscribing for a service in the coverage area of the eNodeB and $\beta_i$ is the $i^{th}$ user subscription weight assigned by the network. Each UE is allocated at least the total amount of its applications target rates if it has any.

The objective function in the optimization problem (\ref{eqn:SC_opt_sub1}) is equivalent to $\sum_{i=1}^{M+N}\beta_i \log(X_i(r_i))$.
%
%
Therefore, optimization problem (\ref{eqn:SC_opt_sub1}) is a convex optimization problem and there exists a unique tractable global optimal solution $r_i^{\text{opt}}$ for each of the $M+N$ users as shown in Corollary (III.1) \cite{Ahmed_Utility3}.
%
\subsubsection{Second-Stage of the Second-Case Optimization Problem}\label{SC-SS-RA}
%
Each one of the $M+N$ UEs allocates optimal rates $r_{ij}^{opt}$ to its applications. Each UE first allocates the application target rate to each of its applications if it is assigned one. It then starts allocating the remaining resources among all the applications based on the application differentiation weight and the type of the application. This optimization problem is solved internally in the UE and can be written for the $i^{th}$ UE as follows:
\begin{equation}\label{eqn:SC_opt_sub2}
\begin{aligned}
& \underset{\textbf{r}_i}{\text{max}}
& & \prod_{j=1}^{L_i}U_{ij}^{\alpha_{ij}}(r_{ij}+c_{ij}) \\
& \text{subject to}
& & \sum_{j=1}^{L_i}(r_{ij}+c_{ij}) \leq r_i^{\text{opt}}\\
& & & r_{ij} \geq 0, \;\;\;\;\; j = 1,2, ...,L_i.
\end{aligned}
\end{equation}
where $\textbf{r}_i =\{r_{i1},r_{i2},...,r_{iL_{i}}\}$, $r_{i}^{\text{opt}}$ is the rate allocated by eNodeB to the $i^{th}$ UE in the first-stage and $c_{ij}$ is same as before.
The objective function of the optimization problem (\ref{eqn:SC_opt_sub2}) is equivalent to $\sum_{j=1}^{L_i}\alpha_{ij} \log(U_{ij}(r_{ij}+c_{ij}))$. Therefore, optimization problem (\ref{eqn:SC_opt_sub2}) is a convex optimization problem and there exists a unique tractable global optimal solution as shown in Corollary (III.2) \cite{Ahmed_Utility3}. Each UE allocates an optimal rate $r_{ij}^{\text{opt}}=r_{ij}+c_{ij}$ to each of its applications.
%
\section{Algorithms}\label{sec:two-stage-alg}
As mentioned before, the RA for the multi-application users with different priorities is achieved in two-stages. In the first-stage, the eNodeB and the UEs collaborate to allocate optimal rates $r_i$ for users as shown in VIP UE Algorithm (\ref{alg:VIP_UE_first-stage}), regular UE Algorithm (\ref{alg:UE_first-stage}) and eNodeB Algorithm (\ref{alg:eNodeB_first-stage}). In the second-stage, the UE internal algorithm allocates applications rates $r_{ij}$ to the UE's applications as shown in the internal UE Algorithm (\ref{alg:internal_UE}).
\subsection{First-Stage RA Algorithm}\label{sec:UE_alloc_alg}
The first-stage of the RA algorithm is presented in this section. The algorithm starts when each UE transmits an initial bid $w_{i}(1)$ to the eNodeB. Additionally, each VIP UE transmits its applications target rates to the eNodeB. The eNodeB checks whether the $\sum_{i=1}^{M}\sum_{i=1}^{L_i}r_{ij}^{t}$ is less or greater than $R$ and sends a flag with this information to each UE. In the case of $\sum_{i=1}^{M}\sum_{i=1}^{L_i}r_{ij}^{t}\geq R$, the regular UEs will not be allocated any of the resources and will not be sending any further bids to the eNodeB.
\begin{algorithm}
\caption{VIP UE Algorithm} \label{alg:VIP_UE_first-stage}
\begin{algorithmic}
\STATE {Send initial bid $w_{i}(1)$ to eNodeB}
\STATE {Send the applications target rates $r_{ij}^{\text{t}}$ to eNodeB}
\LOOP
    \WHILE {Flag $\sum_{i=1}^{M}\sum_{j=1}^{L_i}r_{ij}^{t}\geq R$ from eNodeB}%
	\STATE {Receive shadow price $p(n)$ from eNodeB}
	\IF {STOP from eNodeB} %
	\STATE {Calculate allocated rate $r_{i} ^{\text{opt}}= \frac{w_{i}(n)}{p(n)}$}
			\ELSE
	\STATE {Solve $r_{i}(n) = \arg \underset{r_{i}}\max \Big(\beta_i \log X_i(r_{i}) - p(n)r_{i}\Big)$}
	\STATE {Send new bid $w_{i} (n)= p(n) r_{i}(n)$ to eNodeB}
		\ENDIF
        \ENDWHILE
    \WHILE {Flag $\sum_{i=1}^{M}\sum_{j=1}^{L_i}r_{ij}^{t}\textless R$ from eNodeB}%
	\STATE {Receive shadow price $p(n)$ from eNodeB}
	\IF {STOP from eNodeB} %
	\STATE {Calculate allocated rate $r_{i}^{\text{opt}}= \frac{w_{i}(n)}{p(n)}$}
			\ELSE
	\STATE {Solve $r_{i}(n)=\arg \underset{r_{i}}\max \Big(\beta_i \log X_i(r_{i}) - p(n)(r_{i}+\sum_{j=1}^{L_i}r_{ij}^{t})\Big)$}
	\STATE {Calculate new bid $w_{i}(n)= p(n) (r_{i}(n)+\sum_{j=1}^{L_i}r_{ij}^{t})$}
    \IF {$|w_i(n) - w_i(n-1)| > \Delta w$}
    \STATE {$w_i(n) =w_i(n-1) + \text{sign}(w_i(n) -w_i(n-1))\Delta w(n)$} \\
    \COMMENT {$\Delta w(n) = l_1 e^{-\frac{n}{l_2}}$}
		\ENDIF
    \STATE{Send new bid $w_{i}(n)$ to eNodeB}
		\ENDIF
        \ENDWHILE
\ENDLOOP
\end{algorithmic}
\end{algorithm}
Each VIP UE checks whether the difference between the current received bid and the previous one is less than a threshold $\delta$, if so it exits. Otherwise, the eNodeB calculates the shadow price $p(n)=\frac{\sum_{i=1}^{M}w_{i}(n)}{R}$ and send it to the VIP UEs where it is used to calculate the $i^{th}$ VIP UE rate $r_{i}(n)$ which is the solution of the optimization problem $r_{i}(n)=\arg \underset{r_{i}}\max \Big(\beta_i \log X_i(r_{i})- p(n)r_{i}\Big)$ where $X_i(r_i) = \prod_{j=1}^{L_i}U_{ij}^{\alpha_{ij}}(r_{ij})$. A new bid $w_{i}(n)=p(n) r_{i}(n)$ is then calculated and the VIP UEs check the fluctuation condition as in \cite{Ahmed_Utility2} and send their new bids to the eNodeB. The Algorithm is finalized by the eNodeB. Each VIP UE then calculates its allocated rate $r_{i} ^{\text{opt}}=\frac{w_{i}(n)}{p(n)}$.
\begin{algorithm}
\caption{Regular UE Algorithm} \label{alg:UE_first-stage}
\begin{algorithmic}
\STATE {Send initial bid $w_{i}(1)$ to eNodeB}
\LOOP
	\WHILE {Flag $\sum_{i=1}^{M}\sum_{j=1}^{L_i}r_{ij}^{t}\geq R$ from eNodeB} %
	\STATE {Allocated rate $r_{i} ^{\text{opt}}=0$}
        \ENDWHILE
	\WHILE {Flag $\sum_{i=1}^{M}\sum_{j=1}^{L_i}r_{ij}^{t}\textless R$ from eNodeB} %
    \STATE {Receive shadow price $p(n)$ from eNodeB}
    \IF {STOP from eNodeB} %
	\STATE {Calculate allocated rate $r_{i} ^{\text{opt}}=\frac{w_{i}(n)}{p(n)}$}
			\ELSE
	\STATE {Solve $r_{i}(n) = \arg \underset{r_{i}}\max \Big(\beta_i \log X_i(r_{i}) - p(n)r_{i}\Big)$}
	\STATE {Calculate new bid $w_{i} (n)= p(n) r_{i}(n)$}
    \IF {$|w_i(n) - w_i(n-1)| > \Delta w$}
    \STATE {$w_i(n) =w_i(n-1) + \text{sign}(w_i(n) -w_i(n-1))\Delta w(n)$} \\
    \COMMENT {$\Delta w(n) = l_1 e^{-\frac{n}{l_2}}$}
		\ENDIF
    \STATE{Send new bid $w_{i}(n)$ to eNodeB}
        \ENDIF
        \ENDWHILE
\ENDLOOP
\end{algorithmic}
\end{algorithm}

In the case of $\sum_{i=1}^{M}\sum_{i=1}^{L_i}r_{i}^{t}\textless R$, a flag with this information is sent to each UE by the eNodeB. Each UE checks whether the difference between the current received bid and the previous one is less than a threshold $\delta$, if so it exits. Otherwise, the eNodeB calculates the shadow price $p(n)=\frac{\sum_{i=1}^{M+N}w_{i}(n)}{R}$ and send it to each UE where it is used by the VIP UE to calculate the rate $r_{i}=r_{i}(n)+\sum_{j=1}^{L_i}r_{ij}^{\text{t}}$, $r_i(n)$ is the solution of the optimization problem $r_{i}(n) = \arg \underset{r_{i}}\max \Big(\beta_i \log X_i(r_{i}) - p(n)(r_{i}+\sum_{j=1}^{L_i}r_{ij}^{\text{t}})\Big)$ where $X_i(r_i) = \prod_{j=1}^{L_i}U_{ij}^{\alpha_{ij}}(r_{ij}+c_{ij})$. A new bid $w_{i}(n)= p(n)(r_{i}(n)+\sum_{j=1}^{L_i}r_{ij}^{\text{t}})$ is calculated by the VIP UE. All VIP UEs check the fluctuation condition and send their new bids to the eNodeB. On the other hand, the regular UEs receive $p(n)$ and calculate the rate $r_{i}(n)$ which is the solution of the optimization problem $r_{i}(n) = \arg \underset{r_{i}}\max \Big(\beta_i \log X_i(r_{i}) - p(n)r_{i}\Big)$ where $X_i(r_i) = \prod_{j=1}^{L_i}U_{ij}^{\alpha_{ij}}(r_{ij}+c_{ij})$. A new bid $w_{i}(n)= p(n) r_{i}(n)$ is calculated by the regular UE. All regular UEs check the fluctuation condition and send their new bids to the eNodeB. The Algorithm is finalized by the eNodeB. Each VIP and regular UE then calculates its allocated rate $r_{i} ^{\text{opt}}= \frac{w_{i}(n)}{p(n)}$.
\begin{algorithm}
\caption{eNodeB Algorithm} \label{alg:eNodeB_first-stage}
\begin{algorithmic}
\LOOP
	\STATE {Receive bids $w_{i}(n)$ from UEs}
	\COMMENT{Let $w_{i}(0) = 0\:\:\forall i$}
    \STATE {Receive applications target rates from VIP UEs}
	\WHILE {$\sum_{i=1}^{M}\sum_{j=1}^{L_i}r_{ij}^{t}\geq R$} %
    \STATE {Send flag $\sum_{i=1}^{M}\sum_{j=1}^{L_i}r_{ij}^{t}\geq R$ to all UEs}
    \IF {$|w_{i}(n) -w_{i}(n-1)|<\delta$, $i=\{1,....,M\}$} %
	\STATE {STOP and allocate rates (i.e $r_{i}^{\text{opt}}$ to VIP user $i$)}
		    \ELSE
	\STATE {Calculate $p(n) = \frac{\sum_{i=1}^{M}w_{i}(n)}{R}$, $i=\{1,....,M\}$}
	\STATE {Send new shadow price $p(n)$ to VIP UEs}
        \ENDIF
        \ENDWHILE
	\WHILE {$\sum_{i=1}^{M}\sum_{j=1}^{L_i}r_{ij}^{t}\textless R$} %
    \STATE {Send flag $\sum_{i=1}^{M}\sum_{j=1}^{L_i}r_{ij}^{t}\textless R$ to all UEs}		
    \IF {$|w_{i}(n) -w_{i}(n-1)|<\delta  \:\:\forall i$} %
	\STATE {STOP and allocate rates (i.e $r_{i}^{\text{opt}}$ to user $i$)}
		    \ELSE
    \STATE {Calculate $p(n) = \frac{\sum_{i=1}^{M+N}w_{i}(n)}{R}$}
	\STATE {Send new shadow price $p(n)$ to all UEs}
        \ENDIF
        \ENDWHILE
\ENDLOOP
\end{algorithmic}
\end{algorithm}
\subsection{Second-Stage RA Algorithm}\label{sec:app_alloc_alg}
The second-stage of RA is presented in this section and shown in Algorithm (\ref{alg:internal_UE}) where the rates $r_{ij}$ are allocated internally by the UE to its applications. Each UE uses its allocated rate $r_i^{\text{opt}}$ in the first-stage to solve the optimization problem $\textbf{r}_i =  \arg \underset{\textbf{r}_i}\max \sum_{j=1}^{L_i}(\alpha_{ij}\log U_{ij}(r_{ij}+c_{ij})- p (r_{ij}+c_{ij})) +p r_i^{\text{opt}}$. The rate $r_{ij}^{opt}=r_{ij}+c_{ij}$ is then allocated to the UE's $j^{th}$ application.
\begin{algorithm}
\caption{Internal UE Algorithm}\label{alg:internal_UE}
\begin{algorithmic}
\LOOP
\STATE {Receive $r_i^{\text{opt}}$ from eNodeB Algorithm (\ref{alg:VIP_UE_first-stage}), (\ref{alg:UE_first-stage}) and (\ref{alg:eNodeB_first-stage})}
\STATE {Solve \\$\textbf{r}_i =  \arg \underset{\textbf{r}_i}\max \sum_{j=1}^{L_i}(\alpha_{ij}\log U_{ij}(r_{ij}+c_{ij})- p (r_{ij}+c_{ij})) +p r_i^{\text{opt}}$}\\
\COMMENT {$\textbf{r}_i = \{r_{i1}, r_{i2}, ..., r_{iL_i}\}$}
\STATE {Allocate $r_{ij}^{opt}=r_{ij}+c_{ij}$ to the $j^{th}$ application}
\ENDLOOP
\end{algorithmic}
\end{algorithm}
\section{Simulation Results}\label{sec:sim}
In this section, we consider one eNodeB with four UEs in its coverage area subscribing for a mobile service. The first and second UEs are VIP UEs and the third and fourth UEs are regular UEs. Each one of the four UEs is running two applications simultaneously. The first application is a real-time application whereas the second application is a delay-tolerant application.

We applied algorithm (\ref{alg:VIP_UE_first-stage}), (\ref{alg:UE_first-stage}), (\ref{alg:eNodeB_first-stage}) and (\ref{alg:internal_UE}) in C++ to the UEs functions. The simulation results showed convergence to the optimal global point in the two stages of the algorithm. We present the simulation results for the four users. The first UE is a VIP UE, we use a normalized sigmoidal-like utility function that is expressed by equation (\ref{eqn:sigmoid}) to represent its first application with $a = 3$, $b=20$ which is an approximation to a step function at rate $r =20$ and we set $r_{11}^{t}=20$. Additionally, for the second application of the first user (VIP user) we use a logarithmic function that is expressed by equation (\ref{eqn:log}) with $k = 3$ which is an approximation of a delay-tolerant application. The second user is a VIP user, we use a normalized sigmoidal-like utility function to represent its first application with $a = 1$, $b=30$ and we set $r_{21}^{t}=30$. Additionally, for the second application of the second user (VIP user) we use a logarithmic function with $k = 0.5$ to represent its delay tolerant application. The same parameters of the first user are used for the third user's utility functions except that its applications are not assigned applications target rates. Also, the same parameters of the second user are used for the fourth user's utility functions except that its applications are not assigned applications target rates. Furthermore, we set $\beta_i=1$ for all UEs. We use $r_{max}=100$ for all logarithmic functions, $l_1=5$ and $l_2=10$ in the fluctuation decay function of the algorithm and $\delta=10^{-3}$. Let the application weight $\alpha_{ij}$ in the set $\alpha$ corresponds to the $j^{th}$ application of user $i$ where $\alpha$ be  $\alpha = \{\alpha_{11}, \alpha_{12}, \alpha_{21}, \alpha_{22}, \alpha_{31}, \alpha_{32}, \alpha_{41}, \alpha_{42}\}$.

\begin{figure}
\centering
\includegraphics[height=1.5in, width=3.5in]{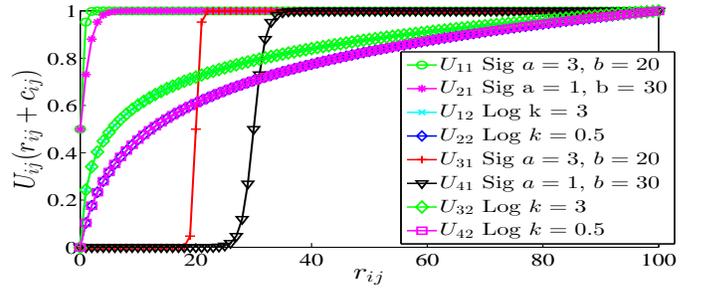}
\caption{The applications utility functions $U_{ij}(r_{ij})$.}
\label{fig:Utility_Uij_Case2}
\end{figure}

\begin{figure}
\centering
\includegraphics[height=1.5in, width=3.5in]{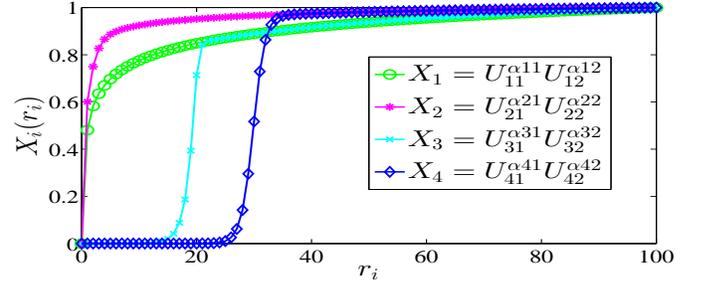}
\caption{The aggregated utility functions $X_i(r_i)$ of the $i^{th}$ user.}
\label{fig:Utility_Xi_Case2}
\end{figure}

Figure \ref{fig:Utility_Uij_Case2} shows eight applications utility functions corresponding to the four UEs. The real-time applications of the VIP UEs are assigned applications target rates, this explains their shifted utility functions by the amount of $r_{ij}^{t}$ in Figure \ref{fig:Utility_Uij_Case2}.
The other applications do not have applications target rates ($c_{ij}=0$ for each one). Figure \ref{fig:Utility_Xi_Case2} shows the aggregated utilities $X_i(r_i)$ for each user.
%
%
\subsection{Convergence Dynamics for $5 \le R \le 200$}
\begin{figure}
\centering
\includegraphics[height=1.5in, width=3.5in]{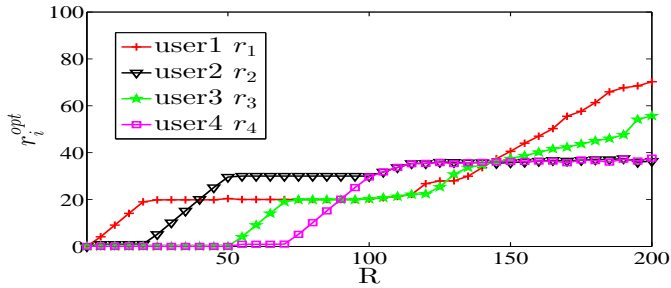}
\caption{The users optimal rates $r_i^{opt}$ for different values of $R$.}
\label{fig:ri_versus_R}
\end{figure}

In the following simulations, we set $\alpha=\{0.5,0.5,0.9,0.1,0.5,0.5,0.9,0.1\}$ and the eNodeB available resources $R$ takes values between $5$ and $200$ with step of $5$. In Figure \ref{fig:ri_versus_R}, we show the four users optimal rates $r_i^{opt}$ with different eNodeB resources $R$. This represents the solution of optimization problem (\ref{eqn:opt_sub1}) when $R\leq50$ and optimization problem (\ref{eqn:SC_opt_sub1}) when $R>50$, using the first-stage of the algorithm, where $50$ is the total applications target rates for the the two VIP users. Figure \ref{fig:ri_versus_R} shows that when $R\leq 50$ the regular UEs are not allocated any of the eNodeB resources. 
Furthermore, when $R>50$ each VIP user is first allocated its total applications target rates and the remaining resources are then allocated to all users based on the proportional fairness approach.

\begin{figure}
\centering
\includegraphics[height=1.5in, width=3.5in]{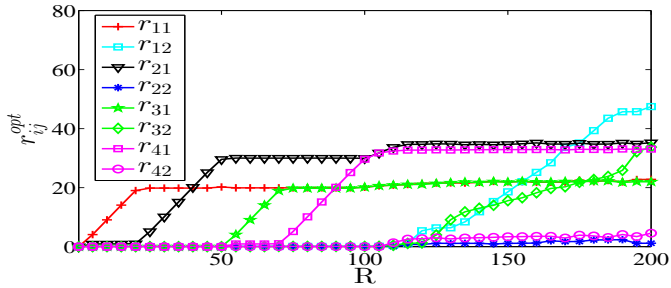}
\caption{The applications optimal rates $r_{ij}^{opt}$ for different values of $R$.}
\label{fig:rij_versus_R}
\end{figure}

In Figure \ref{fig:rij_versus_R}, we show the final optimal  applications rates $r_{ij}^{opt}$ for the four users with different eNodeB resources $R$. This is the solution of optimization problem (\ref{eqn:opt_sub2}) when $R\leq50$ and the solution of (\ref{eqn:SC_opt_sub2}) when $R>50$ using the user internal algorithm. The figure shows that when $R\leq50$, the real-time applications are given priority over the delay tolerant applications when allocating rates by each VIP UE to its applications whereas when $R>50$, the VIP UEs first allocate the applications target rates to the applications that are assigned ones and then allocate the remaining resources among all applications using proportional fairness approach while giving the priority to the real-time applications. The regular users also give the priority to their real-time applications when allocating resources as shown in the same figure.
\subsection{Rate Allocation Sensitivity to change in $\alpha$}
\begin{figure}
\centering
\includegraphics[height=1.5in, width=3.5in]{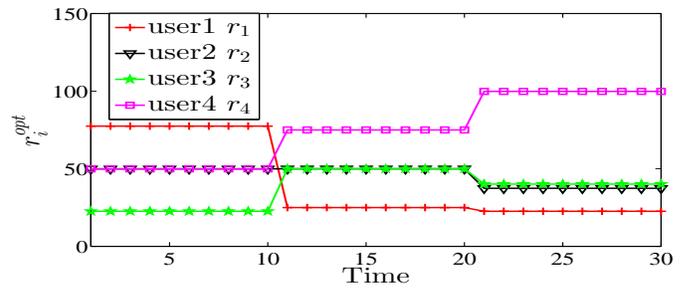}
\caption{The users optimal rates $r_{i}^{opt}$ with the change in users' applications usage percentages $\alpha(t)$.}
\label{fig:DifferentAlphas_Case2}
\end{figure}

In the following simulations, we measure the sensitivity of the change in application weight that is corresponding to the application usage percentage in the UE. We use $R=200$ and the same parameters as before for the four users. The users change their applications usage percentage with time as the following
\begin{equation}
\alpha(t) = \left\{
  \begin{array}{l l}
    \alpha = \{0.1,0.9,0.5,0.5,0.9,0.1,0.5,0.5\};\\
    \;\;\;\;\;\;\;\;\;\;\;\;\;\;\;\;\;\;\;\;\;\;\;\;\;\;\;\;\;\;\;\;\;\;\;for\;\; 0  \le t \le 10\\
    \alpha = \{0.5, 0.5, 0.3, 0.7, 0.2, 0.8, 0.1, 0.9\};\\
    \;\;\;\;\;\;\;\;\;\;\;\;\;\;\;\;\;\;\;\;\;\;\;\;\;\;\;\;\;\;\;\;\;\;\;for\;\; 10  \le t \le 20\\
    \alpha = \{1.0, 0.0, 0.9, 0.1, 0.8, 0.2, 0.1, 0.9\};\\
    \;\;\;\;\;\;\;\;\;\;\;\;\;\;\;\;\;\;\;\;\;\;\;\;\;\;\;\;\;\;\;\;\;\;\;for\;\; 20  \le t \le 30\\
  \end{array} \right.
\end{equation}
Figure \ref{fig:DifferentAlphas_Case2} shows the users optimal rates $r_{i}^{opt}$ with time for the changing usage percentages given by $\alpha(t)$.
\section{Conclusion}\label{sec:conclude}
In this paper, we proposed a novel RA approach to allocate a single eNodeB resources optimally among multi-application UEs with different priority. Two cases of RA optimization problems are considered in our approach. The two cases are based on the total applications target rates of the VIP UEs compared to the eNodeB available resources. A two-stage RA algorithm is presented for each case to allocate the eNodeB resources among users and their applications. 
Different parameters are taken into consideration by our algorithm when allocating resources such as the application type, the application target rate (if the user application has one), the user subscription weight and the application weight. We showed through simulations that our two-stage RA algorithm converges to the optimal rates.
\bibliographystyle{ieeetr}
\bibliography{pubs}
\end{document}